\documentstyle[12pt]{article}
\title{On algebraic classification of quasi-exactly solvable
matrix models}
\author{R.Z.~Zhdanov \\ \small Institute of Mathematics,\\ 
\small 3 Tereshchenkivska Street,
252004 Kiev, Ukraine\thanks{e-mail: rzhdanov@apmat.freenet.kiev.ua}} 
\date{August 17, 1997}
\let\p\partial
\newtheorem{theo}{Theorem}
\begin{document}
\maketitle
\begin{abstract}
We suggest a generalization of the Lie algebraic approach for
constructing quasi-exactly solvable one-dimensional Schr\"odinger
equa\-ti\-ons which is due to Shifman and Turbiner in order to include
into consideration matrix models. This generalization is based on
representations of Lie algebras by first-order matrix differential
operators. We have classified inequivalent representations of the
Lie algebras of the dimension up to three by first-order matrix
differential operators in one variable. Next we describe invariant
finite-dimensional subspaces of the representation spaces of the one-,
two-dimensional Lie algebras and of the algebra $sl(2, {\bf R})$. These
results enable constructing multi-parameter families of first- and
second-order quasi-exactly solvable models. In particular, we have
obtained two classes of quasi-exactly solvable matrix Schr\"odinger
equations.
\end{abstract}

\section*{1. Introduction}

There exists a small number of remarkable Hamiltonians (called
exactly-solvable) whose spectra and corresponding eigenfunctions can
be computed in a purely algebraic way (see, e.g. \cite{per}). However,
the choice of such Hamiltonians is too restricted to meet numerous
requirements coming from different fields of modern quantum physics.
Recently, an intermediate class of Hamiltonians was introduced by
Turbiner \cite{tur} and Ushveridze \cite{ush1} which allow an
algebraic characterization of the part of their spectra. They call
spectral problems of this kind {\em quasi-exactly solvable}.

Quasi-exactly solvable models have an amazingly wide range of
applications in different fields of theoretical physics including
conformal quantum field theories \cite{mor}, solid-state physics
\cite{uly} and Gaudin algebras (an excellent survey on this subject
and an extensive list of references can be found in the monograph by
Ushveridze \cite{ush2}). So it was only natural that there appeared
different approaches to constructing quasi-exactly solvable models,
including the one based on their conditional symmetries (for more
details, see \cite{zhd}). However, for the purposes of this paper the
most appropriate is the Lie-algebraic approach suggested by Shifman
and Turbiner \cite{shi1, shi2} and further developed by
Gonz\'alez-L\'opez, Kamran and Olver \cite{gon1}--\cite{gon3}. That is
why we will give its brief description (further details can be found
in \cite{ush2}).

The Lie-algebraic approach to constructing quasi-exactly solvable
one-dimen\-si\-on\-al stationary Schr\"odinger equations
\begin{equation}
\label{sta}
\left( -{d^2\over dx^2} + V(x)\right)\psi(x)=\lambda \psi(x)
\end{equation}
relies heavily upon the properties of representations of the algebra
$sl(2,{\bf R})$
\[
[Q_0,\ Q_{\pm}]=\pm Q_{\pm},\quad [Q_-,\ Q_+]=2Q_0
\]
by first-order differential operators. Namely, the approach in
question utilizes the fact that the representation space of the
algebra $sl(2,{\bf R})$ having the basis elements
\begin{equation}
\label{lvf}
Q_-={d\over dx},\quad Q_0=x{d\over dx} - \frac{n}{2},
\quad Q_+=x^2{d\over dx} - nx,
\end{equation}
where $n$ is an arbitrary natural number, has an $(n+1)$-dimensional
invariant subspace. Its basis is formed by the polynomials in $x$ of the
order not higher than $n$. Due to this fact, any bilinear combination of
the operators (\ref{lvf}) with constant coefficients yields a
quasi-exactly solvable Hamiltonian ${\cal H}$ such that the
equation ${\cal H}\psi=\lambda \psi$ can always be reduced to the
form (\ref{sta}) with the help of a transformation
\[
\psi(x)\to F(x)\tilde\psi(f(x)).
\]

Note that the above described procedure does not guarantee that
eigenfunctions of thus constructed quasi-exactly solvable Hamiltonians
will be square-integrable. What can be done within this approach is to
reduce a \lq differential\rq\, eigenvalue problem to a matrix one. The
matter of a square integrability as well as other analytical properties
of solutions obtained are to be investigated separately by independent
methods (see, e.g.  \cite{olv93}).

Recently, a number of papers devoted to constructing matrix
quasi-exact\-ly solvable models has been published
\cite{tur1}--\cite{rod}. These papers use the same basic idea which is
to fix a concrete subspace of sufficiently smooth multi-component
functions and then to classify all second-order matrix differential
operators leaving this subspace invariant. Furthermore, the above
subspace is chosen to be the space of all multi-component functions
with polynomial components. Posed in this way, the problem of
constructing matrix quasi-exactly solvable includes as a subproblem a
one of classifying realizations of Lie superalgebras by differential
operators. Being very rich in interesting and important results this
approach, however, contains an evident restriction which does not allow
constructing {\em all} possible quasi-exactly solvable models.  What
we mean is the fact that an invariant subspace to be found is not
necessarily formed by functions having polynomial coefficients.  It is
one of the results of the present paper that there exist principally
different invariant subspaces. Thus there is a necessity for developing
alternative approaches to the problem in question that do not require
fixing {\em a priori} an invariant subspace.

Our initial motivation for studying matrix quasi-exactly solvable
problems was to extend the list of exactly solvable Dirac equations of
an electron via separation of variables. To this end we have suggested
in \cite{zh97} a method for constructing matrix quasi-exactly solvable
models based on a direct generalization of the Lie-algebraic approach
for the case of multi-component wave functions. However, it turns out
that the above method is universal enough to be applied for obtaining
second-order quasi-exactly solvable models as well, including the
Schr\"odinger equations with matrix potentials.

Following \cite{zh97} we extend the class to which should belong basis
elements of a Lie algebra under study (say, of the algebra $sl(2, {\bf
  R})$). We define this class as the set of matrix differential
operators
\begin{equation}
\label{matrix}
Q=\xi(x){d\over dx} + \eta(x),
\end{equation}
where $\xi(x)$ is a smooth real-valued function,\ $\eta(x)$ is a
smooth complex-valued $r\times r$ matrix function, and denote it as
${\cal M}$.  The class ${\cal M}$ is closed with respect to the binary
operation
\[
\{Q_1,Q_2\}\to Q_1Q_2-Q_2Q_1\stackrel{\rm def}{=}[Q_1,\ Q_2]
\]
and, consequently, form the infinite-dimensional Lie algebra.

We will classify inequivalent representations of low dimension ($d\le
3$) Lie algebras by operators belonging to ${\cal M}$. Next, we will
study the additional constraints on the form of basis operators of the
one- and two-dimensional Lie algebras imposed by the requirement that
their representation spaces contain finite-dimensional invariant
subspaces. These results will be used to obtain an exhaustive
description of inequivalent representations of the algebra $sl(2, {\bf
R})$ by matrix differential operators (\ref{matrix}) with $r=2$.
Composing linear and bilinear combinations of basis elements of $sl(2,
{\bf R})$ with constant matrix coefficients will yield multi-parameter
first- and second-order quasi-exactly solvable matrix models. 

\section*{2. Classification of representations of low
              dimension Lie algebras}

Since our aim is to get a quasi-exactly solvable model, we have to
impose an additional restriction on the choice of the basis elements of
the Lie algebras to be considered below. Namely, it is supposed that
there exists at least one basis element such that the coefficient of
${d\over dx}$ does not vanish identically. This constraint is required
to avoid purely matrix representations which are useless in context of
quasi-exactly solvable models.

Consider a first-order differential operator $Q=\xi(x)\p_x + \eta(x)$
with $\xi\not\equiv 0$. Note that hereafter we denote ${d\over dx}$ as
$\p_x$. Let the function $f(x)$ be defined by the relation
\[
f(x)=\int\limits^{x}_a{dy\over \xi(y)},\quad a\in {\bf R}
\]
and the matrix $r\times r$ function $F(x)$ be a solution of system of
ordinary differential equations
\[
\xi(x){d F(x)\over dx} + \eta(x) F(x)=0
\]
with ${\rm det}\, F(x)\ne 0$. Then the equivalence transformation
\[
Q\to \tilde Q = \left(F(x)\right)^{-1}\, Q F(x)
\]
with a subsequent change of the dependent variable
\[
\tilde x = f(x)
\]
reduce the operator $Q$ to become $\tilde Q=\p_{\tilde x}$.
Consequently, any one-dimensional Lie algebra of first-order matrix
differential operators $Q=\xi(x)\p_x + \eta(x)$ with $\xi\not\equiv 0$
is equivalent to the algebra $\langle \p_x \rangle$.

Abstract Lie algebras of dimension up to five have been classified
by Mubarakzyanov in \cite{mub}. Below we give the lists of (non-zero)
commutation relations which determine inequivalent Lie algebras of the
dimension up to three. Note that the algebras which are direct sums of
lower dimension Lie algebras are skipped from the lists.
\begin{eqnarray*}
L_{2,1}&:&[Q_1,\ Q_2]=Q_1;\\
L_{3,1}&:&[Q_2,\ Q_3]=Q_1;\\
L_{3,2}&:&[Q_1,\ Q_3]=Q_1,\quad [Q_2,\ Q_3]=Q_1+Q_2;\\
L_{3,3}&:&[Q_1,\ Q_3]=Q_1,\quad [Q_2,\ Q_3]=Q_2;\\
L_{3,4}&:&[Q_1,\ Q_3]=Q_1,\quad [Q_2,\ Q_3]=-Q_2;\\
L_{3,5}&:&[Q_1,\ Q_3]=Q_1,\quad [Q_2,\ Q_3]=aQ_2,\quad (0<|a|<1);\\
L_{3,6}&:&[Q_1,\ Q_3]=-Q_2,\quad [Q_2,\ Q_3]=Q_1;\\
L_{3,7}&:&[Q_1,\ Q_3]=aQ_1-Q_2,\quad [Q_2,\ Q_3]=Q_1+aQ_2,\quad (a>0);\\
L_{3,8}&:&[Q_1,\ Q_2]=Q_1,\quad [Q_1,\ Q_3]=2Q_2,\quad [Q_2,\ Q_3]=Q_3;\\
L_{3,9}&:&[Q_1,\ Q_2]=Q_3,\quad [Q_2,\ Q_3]=Q_1,\quad
[Q_3,\ Q_1]=Q_2.
\end{eqnarray*}
Here $a$ is a real parameter, the symbol $L_{n,m}$ stands
for a Lie algebra of the dimension $n$ numbered by $m$.

Thus there exists only one two-dimensional Lie algebra
$L_{2,1}=\langle Q_1, Q_2\rangle$ which is not a
direct sum of one-dimensional Lie algebras.

If in operator $Q_1=\xi(x)\p_x+\eta(x)$ the coefficient $\xi$ is not
identically zero, then using equivalence transformations defined at the
beginning of this section we can reduce it to the form $Q_1=\p_{\tilde
x}$. Inserting $Q_1=\p_{\tilde x}$, $Q_2=\tilde\xi(\tilde x)\p_{\tilde
x}+ \tilde\eta(\tilde x)$ into the commutation relation $[Q_1,\
Q_2]=Q_1$ and equating the coefficients of the powers of the operator
$\p_{\tilde x}$ yield system of ordinary differential equations for
$\tilde\xi(\tilde x),\ \tilde\eta(\tilde x)$
\[
{d\tilde\xi\over d\tilde x}=1,\quad {d\tilde\eta\over d\tilde x}=0.
\]
Hence we obtain $\tilde\xi=\tilde x+C_1,\ \tilde\eta=A$, where $C_1\in
{\bf R}$ is an arbitrary constant and $A$ is an arbitrary constant
$r\times r$ matrix. Shifting when necessary the variable $\tilde x$ by a
constant $C_1$ we may put $C_1=0$ and thus get $Q_2=\tilde x\p_{\tilde
x} + A$.

If the operator $Q_1$ has the form $\eta(x)$, then by convention the
coefficient of $\p_{x}$ of the operator $Q_2$ does not vanish
identically. Consequently, there exists an equivalence transformation
reducing the latter to the form $Q_2=\p_{\tilde x}$. Substituting
$Q_1=\tilde\eta(\tilde x),\ Q_2=\p_{\tilde x}$ into the commutation
relation of the algebra $L_{2,1}$ and equating the coefficients of the
powers of the operator $\p_{x}$ we obtain the following equations for
$\tilde\eta(x)$:
\[
{d\tilde\eta \over d\tilde x}=-\tilde\eta,
\]
whence
\[
\tilde\eta(\tilde x)=A{\rm e}^{-\tilde x}.
\]
Here $A$ is an arbitrary $r\times r$ constant matrix.

Summing up, we conclude that the two realizations of the algebra
$L_{2,1}$
\begin{eqnarray}
(1)&& Q_1=A{\rm e}^{-x},\quad Q_2=\p_x;\label{b3}\\
(2)&& Q_1=\p_x,\quad Q_2=x\p_x + A\label{b4}
\end{eqnarray}
exhaust the set of all possible inequivalent representations of the
algebra in question within the class of matrix differential operators
${\cal M}$.

In a similar way we have obtained complete lists of inequivalent
representations of the three- and four-dimensional Lie algebras within
the class ${\cal M}$ which are given below.
\begin{eqnarray*}
L_{3,1}:&& Q_1=A,\quad Q_2=\p_x,\quad Q_3=\epsilon\p_x +
Ax+B,\\
&&[A,\ B]=0;\\[3mm]
L_{3,2}:&& Q_1=A{\rm e}^{-x},\quad Q_2=\epsilon{\rm e}^{-x}\p_x +
(B-Ax){\rm e}^{-x},\quad Q_3=\p_x,\\
&&[A,\ B]=-\epsilon A;\\[3mm]
L_{3,3}:&& Q_1={\rm e}^{-x}(\epsilon\p_x + A),\quad
Q_2={\rm e}^{-x}(\alpha\p_x + B),\quad Q_3=\p_x,\\
&&[A,\ B]=\epsilon B -\alpha A;\\[3mm]
L_{3,4}:&&\\
(1)&& Q_1=A{\rm e}^{-x},\quad Q_2={\rm e}^{x}(\epsilon\p_x +
B),\quad Q_3=\p_x,\\
&&[A,\ B]=-\epsilon A;\\
(2)&& Q_1=\p_x,\quad Q_2=A,\quad Q_3=x\p_x + B,\\
&&[A,\ B]=-A;\\[3mm]
L_{3,5}:&&\\
(1)&& Q_1=A{\rm e}^{-x},\quad Q_2={\rm e}^{-ax}(\epsilon\p_x +
B),\quad Q_3=\p_x,\\
&&[A,\ B]=-\epsilon A;\\
(2)&& Q_1={\rm e}^{-x}(\p_x + A),\quad
Q_2=B{\rm e}^{-ax},\quad Q_3=\p_x,\\
&&[A,\ B]=\epsilon aB;\\[3mm]
L_{3,6}:&& Q_1=A\cos x + B\sin x,\quad Q_2=B\cos x - A\sin x,\quad
Q_3=\p_x,\\
&&[A,\ B]=0;\\[3mm]
L_{3,7}:&& Q_1={\rm e}^{-ax}(A\cos x + B\sin x),\quad
Q_2={\rm e}^{-ax}(B\cos x - A\sin x),\\
&&Q_3=\p_x,\\
&&[A,\ B]=0;\\[3mm]
L_{3,8}:&& Q_1=\p_x,\quad Q_2=x\p_x + A,\quad
Q_3=x^2\p_x + 2Ax + B\\
&&[A,\ B]=B;\\[3mm]
L_{3,9}:&& {\rm No\ representations}. 
\end{eqnarray*}
In the above formulae $\alpha$ is an arbitrary constant,\
$\epsilon=0,1$, and  $A$, $B$ are $r\times r$ constant matrices.

\section*{3. Quasi-exactly solvable matrix models}

As a second step of an implementation of the Lie-algebraic approach to
constructing matrix quasi-exactly solvable models we have to pick out
from the whole set of realizations of Lie algebras listed in the
previous section those having finite-dimensional invariant subspaces.

Consider first the one-dimensional Lie algebra $\langle \p_x\rangle$. A
space with basis vectors ${\bf f}_1(x),\ldots, {\bf f}_n(x)$ is
invariant with respect to the action of the operator $\p_x$ if there
exist complex constants $\Lambda_{ij}$ such that
\[
{d{\bf f}_i(x)\over dx}= \sum\limits_{j=1}^n\,
\Lambda_{ij}{\bf f}_j(x)
\]
for all $i=1,\ldots,n$. Solving this system of ordinary differential
equations yields the following expressions for unknown vector-functions
${\bf f}_i$:
\[
{\bf f}_i(x)=\sum_{j=1}^r\, \sum_{k=1}^n\,\left({\rm e}^{\Lambda x}
\right)_{ik} C_{kj}{\bf e}_j,
\]
where $\Lambda$ is the constant $n\times n$ matrix having the entries
$\Lambda_{ij}$;\ $C_{kj}$ are arbitrary complex constants; the symbol
$(A)_{ij}$ stands for the $(i,j)$th entry of the matrix $A$ and ${\bf
e}_1,\ldots,{\bf e}_r$ are constant vectors forming an orthonormal basis
of the space ${\bf R}^r$.

It follows from the general theory of matrices that the above
formulae can be represented in the following equivalent form
(see, e.g. \cite{gan}):
\begin{equation}
\label{c1}
{\bf f}_i(x)=\sum\limits_{j=1}^m\,{\rm e}^{\alpha_jx}\,
\sum\limits_{k=1}^n\,{\cal P}_{ijk}^{[n-m]}(x){\bf e}_k.
\end{equation}
Here $\alpha_1,\ldots,\alpha_m$ are arbitrary complex numbers with
$|\alpha_i|<|\alpha_{i+1}|$,\ the symbol ${\cal P}^{[n-m]}_{ijk}(x)$
stands for an $(n-m)$th degree polynomial in $x$,\ $1\le m\le n$,\
$k=1,\ldots,n$.

As each realization of the low dimension Lie algebras obtained in Sec.II
contains the operator $\p_{x}$, their finite-dimensional invariant
subspaces are necessarily of the form (\ref{c1}). In what follows we
will obtain complete description of finite-dimensional invariant
subspaces of the representation spaces of the representations of the
two-dimensional Lie algebra $L_{2,1}$ given in (\ref{b3}), (\ref{b4}).

First we turn to the case (\ref{b3}). Let us study the restrictions on
the choice of the basis vector-functions (\ref{c1}) imposed by a
requirement that the corresponding vector space $V_n$ is invariant with
respect to the action of the operator $Q_2=A{\rm e}^{-x}$. By
assumption, there exist complex constants $S_{ij}$ such that the
relations
\[
A{\rm e}^{-x}{\bf f}_i=\sum\limits_{j=1}^n\, S_{ij}{\bf
f}_j
\]
hold with $i=1,\ldots,n$. Hence we get
\[
\sum\limits_{j=1}^m\,{\rm e}^{(\alpha_j-1)x}\,\sum\limits_{k=1}^r\,
{\cal P}_{ijk}^{[n-m]}(x)A{\bf e}_k= \sum\limits_{j=1}^m\,{\rm
e}^{\alpha_jx}\,\sum\limits_{k=1}^r\, \tilde{\cal
P}_{ijk}^{[n-m]}(x){\bf e}_k,
\]
where
\[
\tilde{\cal P}_{ijk}^{[n-m]}=\sum\limits_{l=1}^n\,S_{il}
{\cal P}_{ljk}^{[n-m]}.
\]
Comparing the coefficients of ${\rm e}^{\alpha_ix}$ yield that
$\alpha_{i+1}=\alpha_{i}+1,\ i=1,\ldots, m-1$ and furthermore
\begin{equation}
\label{last}
\sum\limits_{k=1}^r\,{\cal P}_{i1k}^{[n-m]}(x)A{\bf e}_k=0
\end{equation}
for all $i=1,\ldots,n$.

Let us choose the new basis of the space ${\bf R}^r$ in such a way
that the first $s$ basis elements ${\bf e}_1,\ldots, {\bf e}_s$
are eigenvectors of the matrix $A$ with zero eigenvalues, namely
\[
A{\bf e}_i={\bf 0},\quad i=1,\ldots,s.
\]
Given this choice of the basis, it follows from (\ref{last}) that ${\cal
P}_{i1k}^{[n-m]}(x)=0,\ k=s+1,\ldots,r$. Hence, we conclude that the
remaining basis vectors ${\bf e}_{s+1}, \ldots,{\bf e}_r$ satisfy the
relations
\[
A{\bf e}_i=\sum\limits_{j=1}^s\, a_{ij}{\bf e}_j,\quad
i=s+1,\ldots, r
\]
with some constant $a_{ij}$.

Thus, the most general $n$-dimensional vector space $V_n$ invariant with
respect to the Lie algebra $\langle\p_x, A{\rm e}^{-x}\rangle$ is
spanned by the vectors
\[
{\bf f}_i(x)={\rm e}^{\alpha x}\sum\limits_{j=2}^m\,{\rm e}^{(j-1)x}\,
\sum\limits_{k=1}^r\,{\cal P}_{ijk}^{[n-m]}(x){\bf e}_k
+{\rm e}^{\alpha x}\sum\limits_{j=1}^s\,{\cal P}_{ij}^{[n-m]}(x)
{\bf e}_j,
\]
where $\alpha$ is an arbitrary complex constant, ${\cal
P}_{ijk}^{[n-N]}(x),\ {\cal P}_{ij}^{[n-N]}(x)$ are arbitrary $(n-N)$th
order polynomials in $x$,\ $i=1,\ldots,n$. And what is more the matrix
$A$ is of the following form:
\[
A=\left(\begin{array}{cc} 0& \tilde A\\
                          0&   0 \end{array}\right),
\]
where $\tilde A$ is an arbitrary constant $s\times (r-s)$ matrix.

Now we turn to the representation (\ref{b4}). It is necessary to
investigate the restrictions on the choice of the basis vector-functions
(\ref{c1}) imposed by a requirement that the corresponding vector space
$V_n$ is invariant with respect to the action of the operator
$Q_2=x\p_x+A$. By assumption, there exist complex constants $S_{ij}$
such that the relations
\[
\left(x{d\over dx}+A\right){\bf f}_i=\sum\limits_{j=1}^n\,
S_{ij}{\bf f}_j
\]
hold with $i=1,\ldots,n$. Inserting the expressions (\ref{c1}) into
these equations and comparing the coefficients of ${\rm
e}^{\alpha_jx}x^k$ we arrive at the conclusion that
$\alpha_1=\cdots=\alpha_m=0$. With this restriction the formulae
(\ref{c1}) give the most general finite-dimensional invariant subspace
of the representation space of the Lie algebra $\langle\p_x,\
x\p_x+A\rangle$
\[
{\bf f}_i(x)=\sum\limits_{k=1}^r\,{\cal P}_{ik}^{[n-1]}(x)
{\bf e}_k,\quad i=1,\ldots,n.
\]

A detailed investigation of finite-dimensional invariant subspaces
admitted by the three- and four-dimensional Lie algebras is in
progress now and will be the topic of our future publications.

In what follows we will construct examples of quasi-exactly solvable
two-component matrix models based on representations of the Lie algebra
$L_{3,8}=sl(2,{\bf R})$. The construction procedure rely upon the
assertion below which is given without proof.
\begin{theo}
The representation space of the algebra $sl(2,{\bf R})$ having the basis
elements
\begin{equation}
\label{basis}
Q_1={\p_x},\quad Q_2=x{\p_x} + {A},\quad
Q_2=x^2{\p_x} + 2x{A} + {B},
\end{equation}
where ${A},{B}$ are constant $2\times 2$ matrices satisfying the
relation $[{A},\ {B}]={B}$, contains a finite dimensional invariant
subspace iff the matrices $A, B$ are of the form
\begin{eqnarray}
&{1)}& A=\left(\begin{array}{cc} -\frac{n}{2} & 0\\
                               0 & -\frac{m}{2}
      \end{array}\right),
\quad B=\left(\begin{array}{cc} 0 & 0\\
                                0 & 0\end{array}\right),\ {\rm or}
\label{rep1}\\[4mm]
&{2)}& A=\left(\begin{array}{cc} -\frac{n}{2} & 0\\
                               0 & \frac{2-n}{2} \end{array}\right),
\quad B=\left(\begin{array}{cc} 0 & 0\\
                                -1 & 0\end{array}\right).
\label{rep2}
\end{eqnarray}
Here $n, m$ are arbitrary natural numbers with $n\ge m$.
\end{theo}

Using the fact that the algebra in question has the Casimir operator
$C=B\p_x+A-A^2$ it is not difficult to become convinced of the fact that
representations of the form (\ref{basis}), (\ref{rep1}) are the direct
sums of two irreducible representations realized on the representation
spaces
\[
{\cal R}_1=\langle \vec e_1, x\vec e_1,\ldots,
x^n\vec e_1\rangle,\quad
{\cal R}_2=\langle \vec e_2, x\vec e_2,\ldots, x^m\vec e_2\rangle,
\]
where $\vec e_1=(1,0)^{\rm T},\ \vec e_2=(0,1)^{\rm T}$.

Representations (\ref{basis}), (\ref{rep2}) are also the direct sums of
two irreducible representations realized on the representation spaces
\begin{eqnarray*}
{\cal R}_1&=&\langle {n}\vec e_1,\ldots,{n}x^j\vec e_1 +
jx^{j-1}\vec e_2,\ldots,{n}x^n\vec e_1+nx^{n-1}\vec e_2\rangle,\\
{\cal R}_2&=&\langle \vec e_2, x\vec e_2,\ldots,
x^{n-2}\vec e_2\rangle.
\end{eqnarray*}

According to the scheme given in the Introduction to get a quasi-exactly
solvable model we have to compose a linear combination of basis elements
of a Lie algebra of differential operators whose representation space
has finite dimensional invariant subspace $V_n$. And what is more,
coefficients of this linear combination are constant matrices of the
corresponding dimension whose action leaves the space $V_n$ invariant.

Consider first the representation (\ref{basis}), (\ref{rep1}) with
$n=m$. According to the above its representation space contains
$2(n+1)$-dimensional invariant subspace $V_{2n+2}$ spanned by the vectors
${\bf e}_1x^j,\ {\bf e}_2x^j,\ j=0,\ldots,n$. A direct verification
shows that $V_{2n+2}$ is invariant with respect to action of any $2\times
2$ matrix. Composing a linear combination of (\ref{basis}), (\ref{rep1})
under $n=m$ with coefficients being arbitrary $2\times 2$ matrices
yields the following quasi-exactly solvable model:
\begin{equation}
\label{qua1}
(A_1+A_2x+A_3x^2){d{\bf u}(x)\over dx} + (A_4 - nxA_3){\bf u}=
\lambda {\bf u}.
\end{equation}
Here $A_1, A_2, A_3, A_4$ are arbitrary constant $2\times 2$ matrices,
${\bf u}(x)$ is a two-component vector-function.

Provided $n=m+1$, the representation (\ref{basis}), (\ref{rep1}) gives
rise to a quasi-exactly solvable matrix model of the form
\begin{equation}
\label{qua2}
(A_1+B_1x+B_2x^2){d{\bf u}(x)\over dx} + (B_3 + B_2 A){\bf u}=
\lambda {\bf u},
\end{equation}
where $A_1$ is an arbitrary $2\times 2$ matrix and $B_1, B_2, B_2$ are
arbitrary upper triangular $2\times 2$ matrices.

At last, if $n>m+1$, then the representation (\ref{basis}), (\ref{rep1})
yields a quasi-exactly solvable matrix model of the form (\ref{qua2}),
where both $A_1$ and $B_1, B_2, B_3$ are arbitrary upper triangular
$2\times 2$ matrices.

A similar analysis shows that the representation (\ref{basis}),
(\ref{rep2}) gives rise to a quasi-exactly solvable model
\begin{equation}
\label{qua3}
(B_1+B_2x+B_3x^2){d{\bf u}(x)\over dx} + (B_4 +
B_3(2x A + B)){\bf u}=
\lambda {\bf u},
\end{equation}
where
\[
B_i=\left(\begin{array}{cc} \lambda_i& b_i\\
                            0&\lambda_i
          \end{array}\right),\quad i=1,\ldots,4
\]
$\lambda_i, b_i$ being arbitrary complex constants.

Needless to say that any linear matrix model obtained from one of the
above quasi-exactly solvable models by a change of variables is in its
turn quasi-exactly solvable. In other words, equations
(\ref{qua1})--(\ref{qua3}) are representatives of equivalence classes of
quasi-exactly solvable models. Other representatives are obtained via
transformation of variables
\begin{equation}
\label{change}
x\to \tilde x =f(x),\quad {\bf u}\to F(x)\tilde {\bf u},
\end{equation}
where $f(x)$ is a smooth function and $F(x)$ is an arbitrary
invertible $2\times 2$ matrix whose entries are smooth functions
of $x$.

In the same way second-order quasi-exactly solvable matrix models are
constructed. In particular, taking a bilinear combination of the
operators (\ref{basis}) with $A, B$ being given either by (\ref{rep1})
or by (\ref{rep2})
\[
{\cal H}=\sum\limits_{i,j=1,\ i\le j}^3\,\alpha_{ij}Q_iQ_j +
\sum\limits_{i=1}^3\,\alpha_jQ_j,
\]
where $\alpha_{ij},\alpha_i$ are arbitrary real constants, yields two
families of quasi-exactly solvable matrix models of the form
\[
{\cal H}{\bf u}=\lambda {\bf u}.
\]
By a suitable transformation (\ref{change}) the latter can be
transformed to become a matrix Schr\"odinger equation
\[
\left(-{d^2\over d\tilde x^2} + V(\tilde x)\right)
\tilde{\bf u}=\lambda\tilde{\bf u}.
\]
Here $V(\tilde x)$ is the $2\times 2$ matrix potential, whose explicit
form depends essentially on the parameters $\alpha_{ij}, \alpha_i$ and
on the integers $n, m$. Let us stress that the matter of
hermiticity of thus obtained matrix potential is by no means clear and
needs special investigation.

With a particular choice of the matrices $A, B$ 
\[
A=\left(\begin{array}{cc} -\frac{n}{2} &0\\
                         0& -\frac{n}{2}\end{array}\right),\quad
B=\left(\begin{array}{cc} 0 &0\\
                         0& 0 \end{array}\right)
\]
the well-known nine-parameter family of scalar quasi-exactly solvable
Schr\"o\-din\-ger equations is obtained \cite{tur}--\cite{ush2}.

\section*{4. Conclusions}

The present paper is aimed primarily at solving the problem of
classification of quasi-exactly matrix models by purely algebraic means.
As a result, we get some classes of systems of first- and second-order
ordinary differential equations such that a problem of finding their
particular solutions reduces to solving {\em matrix} eigenvalue problem.
Now to decide whether a given specific matrix model is solvable within
the framework of an approach expounded above, one has to check whether
it is possible to reduce it with the help of a transformation
(\ref{change}) to one of the canonical forms given in Sec.III. When one
deals with a scalar model this check is being done trivially (see, for
details, \cite{ush2}). However for the case of matrix models it involves
tedious and cumbersome calculations and is by itself rather nontrivial
algebraic problem. As an illustration we will adduce an instructive
example. Consider the following two-component matrix model:
\begin{equation}
\label{exa1}
{\cal H}{\bf u}\equiv \left( ib\sigma_1Q_1 + ia\sigma_2Q_2 +
c_1\sigma_1 + c_2\sigma_2\right){\bf u} = \lambda{\bf u},
\end{equation}
where $a, b, c_1, c_2$ are arbitrary real parameters with
$ab\ne 0$, and $\sigma_1, \sigma_2$ are $2\times 2$
Pauli matrices, and
\[
Q_1=\p_x,\quad Q_2=x\p_x + \left(\begin{array}{cc}1&0\\
                                 0&1\end{array}\right).
\]

This model is quasi-exactly solvable by construction. Making the
change of variables
\begin{eqnarray*}
x&=&\frac{b}{a}\sinh(ay),\\
\vec w(x)&=& (\cosh(ay))^{1/2}\exp\left\{
-\frac{i}{a}\left(c_1\arctan\sinh(ay) + c_2\ln\cosh(ay)
\right)\right\}\\
&&\times\exp\{-i\sigma_3\arctan\sinh(ay)\}\,\vec \psi(y)
\end{eqnarray*}
we reduce (\ref{exa1}) to the Dirac-type equation
\begin{equation}
\label{exa2}
i\sigma_1{d\vec \psi\over dy} + \sigma_2 V(y)\vec \psi(y)=
\lambda \vec\psi,
\end{equation}
where
\[
V(y)={a^2c_2-b^2c_1\sinh(ay)\over ab\cosh(ay)}
\]
is the well-known hyperbolic P\"oschl-Teller potential.  This means
that the Dirac equation (\ref{exa2}) with the hyperbolic
P\"oschl-Teller potential is quasi-exactly solvable.

Thus an application of the obtained results to decide whether a given
model is quasi-exactly solvable requires a considerable experience in
manipulating matrix exponents. Generically, to check which equations of
the form (\ref{exa2}) can be reduced to one of the quasi-exactly
solvable models constructed at the end of the previous section one has
to solve systems of nonlinear algebraic equations.

A technique used in the present paper can be generalized
in order to enable one to classify multi-dimensional
matrix models in a way as it was done for a scalar case
by Gonz\'alez-L\'opez, Kamran and Olver \cite{gon1}--\cite{gon3}.

The last important remark is that the property of quasi-exact
solvability is intimately connected to the conditional symmetry of a
model under study. This fact was for the first time noticed in our paper
\cite{zhd}, where we proved that quasi-exact solvability of stationary
Schr\"odinger equations is in one-to-one correspondence with their
conditional symmetry. As we believe, similar results can be obtained for
matrix quasi-exactly solvable models as well.

\end{document}